\documentclass[format=acmsmall, review=false, screen=true]{acmart}

\usepackage{booktabs} 

\usepackage[ruled]{algorithm2e} 

\acmJournal{TCPS}

\usepackage{xspace}
\usepackage{amssymb}
\usepackage{amsmath}
\usepackage{url}
\usepackage{graphicx}
\usepackage{color}
\usepackage{bbm}
\usepackage{float}
\usepackage{graphicx}
\usepackage{wrapfig}
\usepackage{url}
\usepackage{wrapfig}
\usepackage{hyperref} 
\usepackage{color}
\usepackage{amstext}
\usepackage{enumerate}
\usepackage{amsmath,bm}
\usepackage{fullpage}
\usepackage{mathtools}
\usepackage{subcaption} 
\usepackage[normalem]{ulem}

\usepackage{amsfonts}
\usepackage{booktabs}
\usepackage{siunitx}

\newcommand{\N}{\mathbb{N}}

\usepackage[textsize=tiny, textwidth=1.7cm]{todonotes}

\begin{document}
\title{Towards a Socially Optimal Multi-Modal Routing Platform} 

\author{Chinmaya Samal}
\email{chinmaya.samal.1@vanderbilt.edu}

\author{Liyuan Zheng}
\email{liyuanz8@uw.edu}

\author{Fangzhou Sun}
\email{fangzhou.sun@vanderbilt.edu}
\author{Lillian J. Ratliff}
\email{ratliffl@uw.edu}
\author{Abhishek Dubey}
\email{abhishek.dubey@vanderbilt.edu}



\begin{abstract}
The increasing rate of urbanization has added pressure on the already constrained transportation
networks in our communities. Ride-sharing
platforms such as Uber and Lyft are becoming a more commonplace, particularly in urban environments. While such services may be deemed more convenient than riding public transit due to their on-demand nature, reports show that they do not necessarily
decrease the congestion in major cities. One of the key problems is that typically mobility decision support systems focus on individual utility and react only after congestion appears. 
In this paper, we propose socially considerate multi-modal routing algorithms  that are proactive and consider, via predictions, the shared effect of riders on the overall efficacy of mobility services. We adapt the MATSim simulator framework to incorporate the proposed algorithms present a simulation analysis of a case study in Nashville, Tennessee that assess the effects of our routing models on the traffic congestion for different levels of penetration and adoption of socially considerate routes. 
 Our results indicate that even at a low penetration (social ratio), we are able to achieve an improvement in system-level performance. 

\end{abstract}

%
%


%
%


\maketitle

\section{Introduction}

\begin{figure}
\begin{subfigure}[h]{0.48\linewidth}
\label{fig-nash-traffic}
\includegraphics[width=1.0\columnwidth]{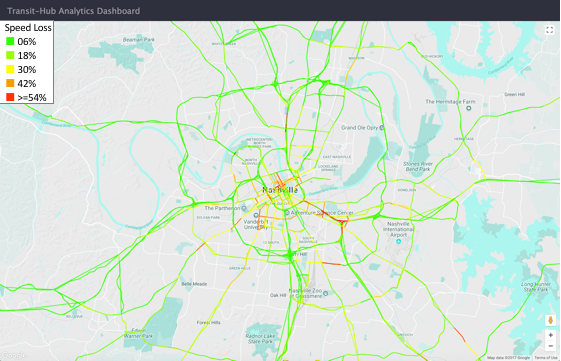}
\caption{Morning (7am-9am) peak hour traffic}
\end{subfigure}
\hfill
\begin{subfigure}[h]{0.48\linewidth}
\includegraphics[width=1.0\columnwidth]{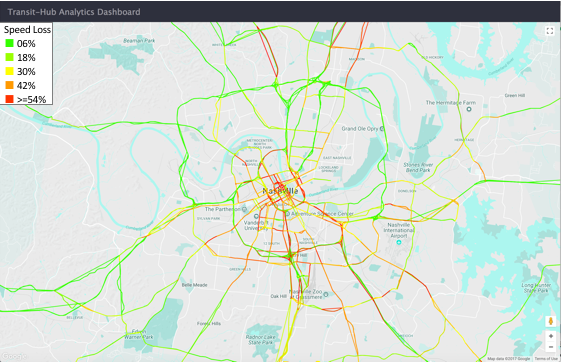}
\caption{Evening (4pm-6pm) peak hour traffic}
\end{subfigure}

\caption{The average speed loss (difference between the free flow speed and real-time speed of a road segment) on all segments in Nashville on September 1, 2017; (a) peak hour traffic at 7am-9am; (b) peak hour traffic at 4pm-6pm.}
\label{fig:traffic-congestion}
\vspace{-0.1in}
\end{figure}

The increasing rate of urbanization has added pressure on the already constrained transportation networks in our communities. For example, a recent estimate
indicates that approximately $100$ people move to the Nashville metropolitan area per day \cite{census2016nashville}.
Commuters predominantly prefer  using their personal vehicle rather than transit options; e.g., according to US Census Bureau's 2013 survey \cite{mckenzie2015drives, census2014biking}, 86\% of all workers commuted to work by private vehicle, either driving alone or carpooling.  This leads to increased congestion,
especially during peak morning and evening travel times.  
Figure~\ref{fig:traffic-congestion} shows the average loss of traffic speed, defined as the relative ratio of decreased speed compared with free flow speed and the free flow speed, on the road during a particular time interval.
The fact that the majority of personal vehicles used by commuters are single occupancy only serves to exacerbate the issue \cite{hu2004summary}.

Technological innovations  have enabled shared mobility options which are increasingly being used by commuters often in lieu of a personal vehicle. To support the demand for such options,  companies are increasingly investing in making shared mobility services readily available to the user on-demand. For example,  ride-sharing apps such as Uber and Lyft are becoming more common place, particularly in urban environments. While such services  may be deemed more convenient than riding public transit due to their on-demand nature, reports show that they do not necessarily decrease the congestion in major cities \cite{ride-share-nactc}.


To fix these problems many cities in the United States have started implementing Transportation Demand Management programs (TDM). For instance, the city of Nashville has the Nashville Complete Trips Transportation Demand Management program \cite{tdm} whose goal is to increase the efficiency of the transportation system and improve the air quality of the region by reducing single occupancy vehicle travel. While these programs are focusing primarily on ad-hoc incentives and introducing new modes of travel (e.g., bike- and car-share options), there is an intrinsic problem that we believe is not being handled. This problem relates to how individuals make a decision of when to travel, which mode of travel to use, and what routes to use and the fact that  decisions are collectively executed on the same constrained, public resource---i.e.~the transportation network. Our hypothesis, which is explored in this paper, is that the current  transportation decision support systems that are commercially available largely focus on short-term planning for individuals (e.g., each user receives information or routing suggestions determined based on their stated preferences) rather than analyzing the short-term societal or system-level impacts of routing decisions across users and then using that analysis in distributing the mobility demand across the dimensions of space, time, and modes of transport.




\subsection{Commercial Mobility Decision Support Solutions} 
To help us illustrate our point, consider that a large number of U.S.~adults use smartphone apps as decision support systems for planning their mobility choices; e.g., Google Maps, Waze, Apple Maps are some of the major trip planners employed by users. In fact, the number of commuters utilizing these services has gone up from 74\% in 2013 to 90\% in 2015 according to one study \cite{pew-research-2015}. 

Users can specify their preferences for certain routes, departure time, and desired means of transport; the mobility apps give a set of itineraries from which the user can select an option that is most closely aligned with their preferences. They also provide users with a traffic heat map showing estimated real-time congestion levels, either predicted via historical data or obtained via real-time samples from other users using the service.  

A prevalent solution with current trip planners is that, at peak hours when routes become more congested, new routes are offered to users based on estimated real-time traffic conditions. While current planners have long-term prediction models that give individual users routes determined to be optimal for that user given estimated current conditions, such a solution is \textit{reactive} in the sense that users change or adjust their travel plans {\it after} congestion occurs. 

\begin{figure}[t]
\vspace{-0.1in}
\begin{center}
\centerline{\includegraphics[width=1.0\columnwidth, height=0.8\columnwidth,keepaspectratio]{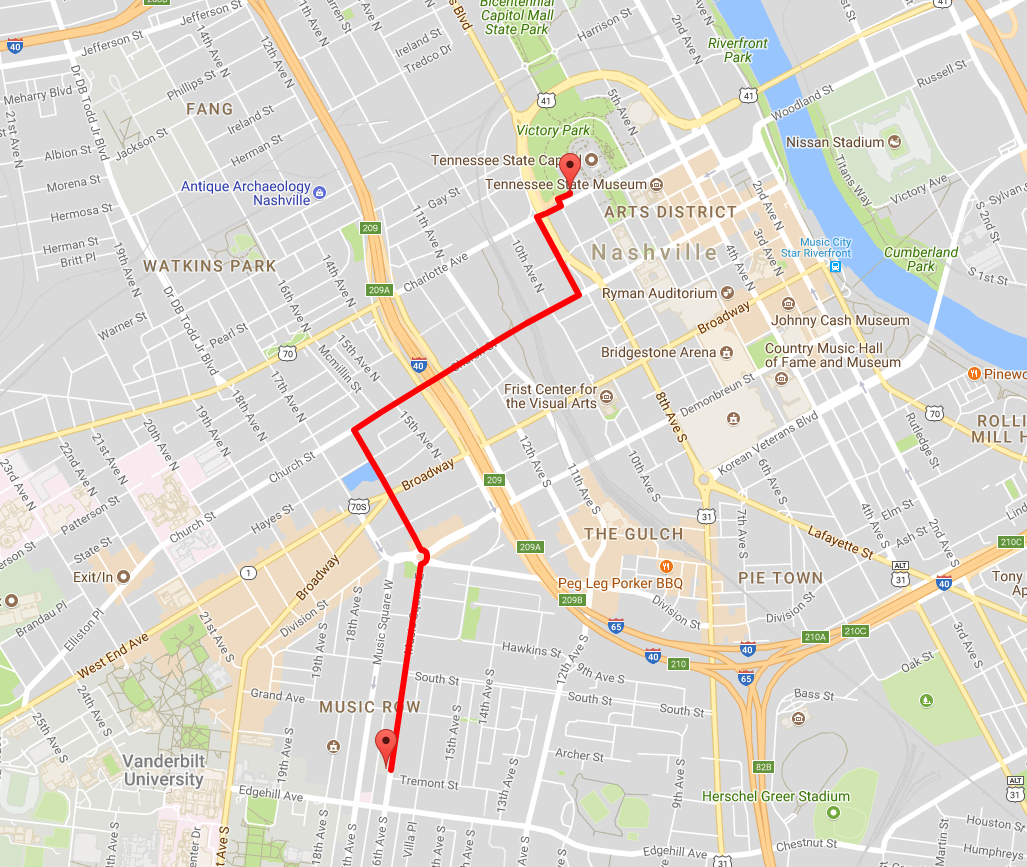}}
\caption{Congestion analysis of Google Maps.  The congestion level of a route is shown when 1000 users make requests to go from Institute for Software Integrated Systems to Music City Central, Nashville, TN by car. Specifically, 200 requests were sent from 5 different machines and the departure time for the requests were between 9AM--9:05AM with requests uniformly distributed across time. 
}
\label{fig-google-heatmap}
\vspace{-0.3in}
\end{center}
\end{figure}

As an example of this phenomena, we used Google Maps\cite{google-maps} to analyze the routes given to users during during peak hours. Specifically, we conducted an experiment to see what happens when 1000 users make a request to go from the same general origin to same general destination in an interval of 5 minutes (requests are distributed uniformly across the 5 minute period). Information on routing algorithms and the congestion model that Google Maps employs is not publicly available. However, by querying their system, we are able to analyze the routes provided via their platform. As shown in Figure \ref{fig-google-heatmap}, one route was provided to all users and, assuming all users take this route, it becomes heavily congested.  We hypothesize that routes produced by Google Maps' are done so in a reactive manner.

 Efficient multi-modal routing has been studied extensively. While routing algorithms are practical and are used by current trip planners, there is a large gap between what is socially optimal (i.e.~at the system-level) and what users choose to do as selfish individuals, as has been studied extensively in the routing game literature. Researchers largely agree that distribution of mobility demand is a potentially viable approach to addressing this gap. Yet, there are some practical challenges: (1) we do not know how many people are participating, (2) there are no simulation tools to study and analyze the policy assumptions (which can be used to design incentives and then improve the solution applications), (3) the dynamic routing problem across multiple modes is a difficult problem. We discuss these problems further in the related work section \ref{sec:relatedwork}.

\subsection{Contributions} 

This paper has following contributions:
\begin{itemize}
\item We propose a User Optimal Multi-Modal Router (UO-MMR) in which the best possible multi-modal paths are given to the user that maximizes their multiple objectives. It improves upon state-of-the-art multi-modal, multi-objective routing algorithms by using switch conditions to provide a feasible multi-modal path in the network and gives results that are user-optimal in the sense that the user objectives are maximized. 

\item Building on the UO-MMR extension, we propose a Social Optimal Multi-Modal Router (SO-MMR) which takes a \emph{proactive} approach to avoid congestion by suggesting route choices that are more socially optimal in the sense that they are considerate of the choices made by other users and work towards improving system-level performance.
\item We present a simulation analysis of the effects of UO-MMR and SO-MMR on the traffic congestion. To this end, we leverage the MATSim simulation tool---an open-source framework for implementing large-scale agent-based transport simulations \cite{horni2016multi}. We assess the effect of the ratio of SO-MMR users to UO-MMR users in the population on system level performance as measured by average travel-time across users\footnote{Of course, other metrics can be used and our framework is agnostic to the particular metric.}.

\end{itemize}

Our results indicate that as the number of users using SO-MMR generated routes increases, the average travel time of all agents in the system decreases and so does the variation in travel times. Our results also indicate that to decrease congestion, SO-MMR increasingly offers transit routes to the users. This re-confirms that multi-modal routing is one of the many important strategies that, if appropriately leveraged, can decrease congestion while also decreasing the travel-time of a significant proportion of users in the system. Not surprisingly, as there is no free lunch, our results also show that some users  do experience longer travel times under the SO-MMR generated routes  as compared to UO-MMR routes; however, the increase in travel time is not significant in the case study we explore. This motivates a future line of research into the fairness of such strategies as well as incentives to encourage participation and route adoption.

\subsection{Paper outline}

In Section~\ref{sec:relatedwork}, we provide a brief overview of the current literature on routing algorithms and routing games as it relates to the work in this paper. Subsequently, in Section~\ref{sec:graph-setup}, we describe the setup of our multi-modal graph-based approach and, in Section~\ref{sec:multi-modal-routing}, we describe the proposed algorithms for multi-modal routers. We detail the simulation framework, execution procedure, and key results findings in Section~\ref{sec-simulation}.  We provide a discussion of the results and future directions in Section~\ref{sec:discussion} and make concluding remarks in Section~\ref{sec:conclusion}.

\section{Related work}
\label{sec:relatedwork}
There are two bodies of work that we draw on to construct our approach to more socially optimal transportation resource consumption. The first is the literature on routing algorithms from which we build on the idea of multi-objective routing in networks with constraints. The second is the literature on routing games. In this latter body of work, there is a long history of studying inefficiencies arising from selfish routing.

\subsection{Routing Algorithms}
\label{related-routing-algos}

Routing is a widely researched topic in large part due to its practical relevance in real-world applications. Much of the work on route planning focuses on finding the shortest path on a directed weighted graph. Dijkstra \cite{dijkstra1959note} and Bellman and Ford \cite{ford1956network, bellman1958routing} proposed some of the first algorithms to solve this problem. Although these algorithms are quite old, they are still fundamental to many route planning algorithms that exist today. 

While these algorithms compute optimal shortest paths, they are too slow to process real-world data sets such as those deriving from large-scale road networks. To address this issue, there are many techniques aimed at speeding up these algorithms. Such techniques often are based on clever heuristics that accelerate the basic shortest paths algorithms by reducing their search space. Bi-directional search \cite{dantzig2016linear, goldberg2005computing}, e.g.,  not only computes the shortest path from the source $s$ to the target $t$, but simultaneously computes the shortest path from $t$ to $s$ on the backward graph. Goal-directed search such as A$^{*}$ \cite{hart1968formal} uses other heuristics  to guide the search.  Goldberg et al.~proposed the \textit{ALT} approach in which they enhance A$^\ast$ by introducing landmarks to compute feasible potential functions using the triangle inequality \cite{goldberg2005computing, goldberg2005computing1}. In other work, contraction techniques are used to speed-up the shortest path computation; e.g., \emph{highway hierarchies} \cite{sanders2005highway, sanders2006engineering} exploits the hierarchical in road networks, while \emph{contraction hierarchies} \cite{geisberger2008contraction} is based on contracting the graph.

Routing in public transportation is more difficult than in road networks due to scheduling which has to be done using largely the same network edges as are used by personal vehicles, thereby introducing uncertainties in the resulting timing plan. There are two approaches to cope with the inherent time-dependency of a public transportation schedule: the time-expanded and the time-dependent approach. The time-expanded model is intuitive and allows more flexibility in constraints while the time-dependent approach has a smaller graph size and, thus, faster query times \cite{pajor2009multi}.

However, in much of this existing work as identified above, only single criteria is used for finding paths. In public transportation networks, multi-criteria optimization is important due to the many potentially conflicting objectives that exist including minimizing travel-time, costs, transfers, etc. This gives rise to a multi-objective search problem, which is an extension of the shortest path problem where link costs are expanded to vectors containing several objectives. This problem has been studied extensively \cite{muller2001pareto, muller2007finding, disser2008multi}.

The multiobjective A$^\ast$ search algorithm finds all of the Pareto-optimal paths (non-dominated paths) in a multi-criteria network \cite{tung1992multicriteria,mandow2005new}. It has been shown that at worst, multiobjective A$^\ast$ search requires an exponential size graph in space and time \cite{mandow2010multiobjective}. Some variants such as relaxed Pareto dominance ($\epsilon$-dominance) are used in practice \cite{perny2008near}. 

We, on the other hand, are interested in the multi-modal setting in which users can choose amongst a combination of modes of transport. Our approach of developing a multi-modal graph (detailed in Section~{sec:graph-setup}) is similar to the work in \cite{pajor2009multi}. To avoid arbitrary modes of transportation at arbitrary points of the network in a multi-modal network, the \emph{label constrained shortest path problem} \cite{barrett2008engineering, barrett2000formal}, where regular languages model reasonable path restrictions, can be used. However, this approach does not take into account the real-time properties of the network and user context. Hence, we propose to use  \emph{switch conditions} \cite{liu2010data} to construct a feasible multi-modal path in a network. Our approach improves upon multi-objective A$^\ast$ \cite{tung1992multicriteria} to include congestion information and route users based on the plans of other users. 

\subsection{ User Choice versus Social Optimum}
\label{subsec:routing-game-literature}

In a related body of work referred to as \emph{static routing games} the notion of inefficiency is well-studied (see, e.g., \cite{van1990tacit,roughgarden2002bad,roughgarden2005selfish,roughgarden2004bounding}). Static routing games are ones in which there is a graph representing a network with a number of commodities (i.e.~source-destination pairs). Populations of users must allocate themselves amongst a finite set of routes (paths connecting sources to destinations) associated with their commodity. Paths are made of up network edges, each of which has an associated congestion-related cost or latency that the users experience if the edge is selected.
Applications range from wireless networks to transportation (see, e.g., \cite{vickrey1969congestion,boyce2005retrospective,youn2008price,sheffi1985urban,correa2008geometric}). There is a large body of work in this area that focuses on characterizing the gap between the selfishly selected user equilibrium (Wardrop equilibrium) \cite{wardrop1952road} and the social optimal solution. This gap is commonly referred to as the \emph{price of anarchy}. 

One avenue of research focuses on how mechanisms such as \emph{tolling}, \emph{uncertainty}, and \emph{altruism} affect equilibrium quality and can serve to reduce the price of anarchy. 
Indeed, the approach of \emph{tolling} places additional costs on edges in order to redistribute or alleviate congestion leading to a narrowing of the gap between the social optimum and user-selected equilibrium. For instance, there are a number of solutions for obtaining the optimal edge pricing \cite{cole2003much,fleischer2004tolls,cole2003pricing,jelinek2014computing}. Similarly, a number of works have examined how users' uncertainty level regarding edge costs or travel information (see, e.g.,  \cite{liu2016effects,wu2017informational,thai2016negative,sekar2017uncertainty}) impacts the price of anarchy.
The price of anarchy under stochastic selfish routing game with risk-averse players has also been studied \cite{nikolova2011stochastic}. 
It has been shown that, counter-intuitively,  under some conditions on structure of edge-level congestion, equilibrium quality under uncertainty is better compared to the full information case \cite{sekar2017uncertainty}. The study of \emph{altruism} is analogous to uncertainty in that it is a property associated with the user. In this case, it is assumed users are (at least partially) willing to suffer (take on additional cost) for the good of society. 
For example, the impact of the degree of cooperation on equilibrium quality has been explored in \cite{azad2010routing} and, in the case when users are partially altruistic, the price of anarchy has been studied in \cite{chen2008altruism,ccolak2016understanding}. 

In large part, the key idea in each of these avenues of work is that users perceive or experience different edge costs than they would under \emph{normal operating conditions}. Yet,
the static approach adopted in much of the routing game literature has some limitations as far as the realism with which it represents the actual
process that gives rise to congestion and increased travel time \cite{chiu:2011aa}. 
The so-called Dynamic Traffic Assignment (DTA) problem, as its name suggests, seeks to address the routing problem in a dynamic setting by determining routes for users over a period of time.
There are largely two directions of research into the DTA problem: non-iterative or iterative. The latter has the benefit of repeatedly solving the problem in order to achieve a more efficient routing solution, however, it cannot be solved in a real-time setting where users are arriving to a system and routes need to be recommended.   
It is our aim to address the routing problem in this context. 

While there are some approaches aimed at closing the gap between the social (or system-level) optimum and the user select behavior \cite{samaranayake2015discrete,qian2012system}.
In line with this body of work, our approach leverages an agent-based simulation environment to assign routing choices that are more socially optimal in the sense that they are considerate of the choices made by other users and work towards improving system level performance which can be metrized, e.g., by average travel time across all users. In addition, we consider multiple modes of travel (e.g., bike, personal or shared vehicle, transit, etc.). Specifically, users are offered route choices aligned with their specified preferences dynamically. Moreover, we assess the level of adoption of socially considerate routes on performance. Such analysis will be the foundation for understanding the level of incentives needed to have an appreciable effect on system level performance.




\section{Problem Setup}
\label{sec:graph-setup}

\subsection{Preliminaries}
Let $G=(V, E)$ be a directed graph, where $V$ is the set of vertices, $E \subseteq V \times V$ the set of edges. We say there is an edge from $u \in V$ to $v \in V$ if and only if $(u, v) \in E$. Note that we use the terms graph and network interchangeably. Each edge $e \in E$ has a set of edge labels $M_e$ that denotes the different modes of transportation allowed on $e$, where $M_e \in \{walk, bike, car, transit\}$. 

A graph is said to be weighted when a numerical label (i.e.~weight) is assigned to each of its
edges. For instance, there might be a cost involved in traveling from a vertex to one of its neighbors, in
which case the weight assigned to the corresponding edge can represent such a cost. All edges in our graphs are weighted by periodic time-dependent travel time functions \cite{pajor2009multi} $f_e : \Pi \rightarrow \N_0$ where $\Pi$ depicts a set of time points or time-period (seconds, minutes or hours of a day). If $f_e$ is constant over $\Pi$, we call $f_e$ as time-independent \cite{pajor2009multi}. In time-dependent graphs, the shortest path depends on the departure time $\tau_s$ of the source node. This might result in shortest paths of different length for different departure times or even a completely different route. A time-query has as input $s \in V$ and a departure time $\tau$. It computes a shortest path tree to every node $u \in V$ when departing at $s$ at time $\tau$.

\subsection{Uni-Modal Graph}

Our multi-modal graph is composed of different uni-modal graphs for each mode of transportation. We briefly discuss different uni-modal graphs relevant in our work. In road networks, nodes model intersections or point of interests (POI) and edges depict street segments. A street segment is often shared by different modes such as $ \{walk, bike, car\}$. In this case, the travel-time for each mode is dependent on other modes with shared street segments. 

In some street segments, the paths for each mode are clearly divided and hence, travel-time for one mode is independent of other modes. In the case of segments that represent roads for vehicles, the segments are shared amongst a variety of vehicle types including cars, trucks, and other heavy duty vehicles.
Travel time on a road segment depends on the congestion. The effect of congestion in road transport, is primarily that the travel time of a congested road segment increases as the traffic load approaches the traffic capacity of the road segment. The travel time increase can be modeled with the BPR (Bureau of Public Roads) \cite{manual1964bureau} formula:
\[T_e(\tau_u^i)=t_{f,e}\left(1+\alpha_e\frac{\mu_e(\tau_u^i)}{c_e}\right)^{\beta_e}\]
where 
\begin{itemize}
\item $\tau_u^i$: time at which the user $i$ departs node $u$ along edge $e$ connecting $u$ to $v$
\item $\mu_e$: function that provides expected volume of vehicles on edge at time $\tau_u^i$
\item $c_e$: capacity of edge $e$ (max number of vehicles that could fit on that edge)
\item $t_{f,e}$: free-flow travel time of edge $e$ (time it would take to traverse edge $e$ if vehicles were moving at the free-flow velocity---i.e.~the speed limit---on that edge) 
\item $\alpha_e$: a constant in the BPR function (usually taken to be $0.15$)
\item $T_e(\mu_e)$: travel time on edge $e$ as a function of the volume $\mu_e$
\item $\beta_e$: a constant in the BPR function (usually taken to be $4$)
\end{itemize}

As can be seen from the formula the travel time will stay at the free flow travel time until the flow is very close to the capacity. The travel time then increases rapidly as the expected volume of vehicles approach the capacity of the edge.

\subsection{Multi-Modal Graph}
\label{sec:multi-modal-graph}

For building a multi-modal graph $G = (V,E)$, we merge the node and edge sets of each individual uni-modal graph. Combining the graph requires us to identify neighbors of each node which has different modes, where the user can switch from one mode to another. Such a node is called \textit{switch node}. These switch nodes are candidate nodes where we \textit{merge} and \textit{link} the uni-modal graphs. The act of merging unites two different uni-modal graphs and the act of linking inserts link edges to connect the nodes having different modes. We only link nodes that are no more than distance $\delta$ apart, a parameter determined by each switch node. Associated with each switch edge are list of pre-conditions, that needs to be satisfied to traverse that edge. Such conditions are called the \textit{switch condition module}. The switch condition module  depends on the properties of the switch nodes, switch edge connecting the switch nodes, user profile and time.
Specifically, define $\text{SCM}(e, U_i, \tau_u^i)$, where $e=(u,v)$ is the switch edge with $u$ and $v$ as the switch nodes, $U_i$ is the profile of user $i$ and $\tau_u^i$ is the departure time at node $u$. Some of the conditions that are used in the switch condition module include the following:
\begin{itemize}
\item Does the user have a mode with it (e.g., car or bike)?
\item If yes, does the user need to park its mode?
\item If not, can the new mode store the current mode (e.g., bike rack on bus)?
\item Does the user have mode rental options (e.g., car- or bike-sharing)?
\item What is the associated cost for switching modes? 
\item Is the cost less than a fixed amount (e.g., bound on the user's available capital)?
\item How much time does it take to switch modes?
\item Are their physical restrictions such as turn movements?
\item Are their additional cost such as toll payments or parking? 
\end{itemize}

To determine the nearest neighbors, we use $k$-dimensional trees (\emph{$k$-d trees}) \cite{bentley1975multidimensional}, a data structure specifically designed for geometric search algorithms. The idea is to generalize a binary search tree to $k$ dimensions. Queries of $k$-dimensional points can be answered in average logarithmic time.

\section{Multi-Modal Routing}
\label{sec:multi-modal-routing}

In this section, we describe our approach to UO-MMR, an improvement of existing multi-modal routing algorithms, that determines which paths should be given to users provided multiple objectives. In addition, we describe our SO-MMR which builds on UO-MMR to offer more \emph{socially considerate} route suggestions to users aimed at reducing congestion in the overall network, thereby improving system performance. 

\subsection{User Optimal Multi-Modal Router (UO-MMR)}

In UO-MMR, the best possible multi-modal paths that maximizes multiple objectives are given to the user. 
UO-MMR is based on the Multi-Objective A$^{\ast}$ algorithm. We use UO-MMR for routing on our multi-modal graph described in Section~\ref{sec:multi-modal-graph}. To find feasible and meaningful multi-modal paths in a network, we are using Algorithm \ref{algo-outgoing-uo} to get out-going edges which satisfy the set of conditions in the switch condition module.

\begin{algorithm}[htbp]
	\KwData{A multi-modal graph $G = (V,E), u \in V$, $\tau_u^i$ = departure time of user $i$ at node $u$ , ($U_i$) = profile of user $i$.}
	
    \KwResult{Set of Links}
    
    \Begin{
    
    	Initialize OutgoingEdgeList
    	
        \ForEach{$v \in u.neighbors$}{
        		e $\longleftarrow$ Edge from $u$ to $v$\;
               		\If{$\mathrm{SCM}(e, U_i, \tau_u^i)$ is True}{
        				OutgoingEdgeList.append(e) \;
      				}
         }

     }

    \Return{OutgoingEdgeList}
	\caption{Get Outgoing Edges UO-MMR}
    \label{algo-outgoing-uo}
\end{algorithm}

\begin{algorithm}[htbp]
	\KwData{{A multi-modal graph $G = (V,E), e = (u,v) \in E$, $P_i$ = Path of user $i$}}
    \KwResult{Congestion Model of links present in $P_i$ is updated}
    \Begin{
    	
        \ForEach{$e=(u,v) \in P_i $}{
                    $\tau_u^i \longleftarrow u.departuretime$\;
                    $\tau_v^i \longleftarrow v.departuretime$\;
                    
                    CongestionModel $\longleftarrow$ $e.intervaltree$\;
                    
                    \If{$e.mode$ == car}{
        				CongestionModel.addInterval($\tau_u^i$, $\tau_v^i$, 1)\;
      				
                    }\uElseIf{$e.mode$ == bus}{
                          e.mode.capacity++\;
                    }
       }

     }

	\caption{Add User Plan}
    \label{algo-add-user-plan}
\end{algorithm}

\begin{algorithm}[htbp]
	\KwData{A multi-modal graph $G = (V,E), e = (u,v) \in E$, $\tau_u^i$ = departure time of user $i$ at node $u$}
	\KwResult{congestion level $\in [0,1]$)}
    
    \Begin{
    	
    	congestion = 0\;
        CongestionModel $\longleftarrow$ $e.intervaltree$\;
        OverlappingIntervals = CongestionModel($\tau_u^i$);
        
        \ForEach{interval $\in$ OverlappingIntervals}{
                    \If{interval.min > $\tau_u^i$}{
        				congestion = congestion + interval.value;
      				}
         }
         
         ratio = $\frac{congestion}{e.capacity}$
       
     }
    
    \Return{ratio}
	\caption{Get Predicted Congestion Level}
    \label{algo-get-congestion-level}
\end{algorithm}

\begin{algorithm}[htbp]
	\KwData{A multi-modal graph $G = (V,E), e = (u,v) \in E$, $\tau_u^i$ = departure time of user $i$ at node $u$ , ($U_i$) = profile of user $i$ and the social ratio $\alpha\in [0,1]$}
	
    \KwResult{Set of Links}
    
    \Begin{
    
    	Initialize OutgoingEdgeList
    	
        \ForEach{$v \in u.neighbors$}{
        		$\delta$ = Get Predicted Congestion Level(e, $\tau_u^i$)\;
        		$e \longleftarrow$ Edge from $u$ to $v$\;
               		\If{$\mathrm{SCM}(e, U_i, \tau_u^i)$ is True \textbf{and} \textbf{$\delta < \alpha \times e.capacity$}}{
        				
                        \If{e.mode == bus \textbf{and} e.mode.capacity is full}{
                        		\textbf{continue};
                        }
                        
                        OutgoingEdgeList.append(e) \;
      				}
         }  	
     }

    \Return{OutgoingEdgeList}
	\caption{Get Outgoing Edges SO-MMR}
    \label{algo-outgoing-so}
\end{algorithm}

\subsection{Social Optimal Multi-Modal Router (SO-MMR)}
Much like UO-MMR, SO-MMR returns the set of best possible multi-modal paths that maximize multiple objectives to the user. Extending beyond UO-MMR, SO-MMR takes a proactive approach to avoiding congestion by suggesting route choices that are more socially optimal in the sense that they are considerate of the choices made by other users and work towards improving system level performance. Unlike UO-MMR, SO-MMR takes into account the route plan given to the user and updates the congestion model of all links present in the route plan. 


To add a user's plan, we take all the links given in the route plan and the departure time present in nodes connecting each link and add congestion to each link as shown in Algorithm \ref{algo-add-user-plan}. The amount of congestion we add depends on the mode the user is using to traverse a link. For a link, if a user is in a car, we increment the congestion level in the time interval the car traverses the link, if it does at all.  We note that buses have a physical capacity limiting the number of users that can be on a bus at a given time. In addition, for a link, we update the congestion according the the bus schedule and delays incurred in the simulation---that is, in the time interval that a bus traverses a link we update the congestion on that link accordingly. The difference between cars and a bus in terms of the congestion model is that cars can be associated with a user's plan whereas buses have a schedule independent of any particular user. 
In our graph, we have assumed that the schedule of the bus is fixed. For $walk$, we assume that adding a user will have negligible effect on congestion on road and hence, we do not update the congestion.

As shown in the Algorithm \ref{algo-get-congestion-level}, to get the predicted time dependent congestion level of a link $e=(u,v)$ at the departure time $\tau_u^i$ of user $i$, we use \emph{interval trees} \cite{cormen1990introduction} to get overlapping intervals and look for intervals $[a,b]$ whose minimum value $a$ is greater than the given time $\tau_u^i$.  The predicted congestion level varies from zero to one, where zero means that there is no congestion and one implies that the link is fully congested. 

Finally, we update the Algorithm \ref{algo-outgoing-uo} to also take into account congestion of links and the capacity of buses (or any transit vehicle). The updated algorithm is shown in Algorithm \ref{algo-outgoing-so} in which we introduce 
 the 
\textit{social ratio} $\alpha\in [0,1]$ defined to be the ratio of the total population using SO-MMR generated routes---that is, 
if $P$ is the total population size, then the population of SO-MMR users is $Y$ with cardinality $|Y|=\alpha P$ and the population of UO-MMR users is $X$ with cardinality $|X|=(1-\alpha)P$. As indicated in the Algorithm \ref{algo-outgoing-uo}, congestion should be less than $\alpha \times e.capacity$. 


Hence, to develop SO-MMR router, we have updated the UO-MMR router (i)  to add a user's plan to the graph as per  Algorithm \ref{algo-add-user-plan}, (ii) obtain outgoing edges via Algorithm \ref{algo-outgoing-so}, and (iii) update link costs using predicted congestion obtained via Algorithm \ref{algo-get-congestion-level}.

\section{Simulation}
\label{sec-simulation}

\subsection{MATSim Implementation}

We use MATSim \cite{horni2016multi} for our simulation. MATSim supports implementing large-scale agent-based transport simulations and is based on iterative dynamic traffic assignment. That is, every agent repeatedly optimizes its daily activity schedule while in competition for space-time slots with all other agents on the transportation infrastructure. Every agent possesses a memory containing a  fixed number of day plans, where each plan is composed of a daily activity chain and an associated score. The score can be interpreted as an econometric utility. 

While MATSim is capable of iteratively finding \emph{best} routes for users, we are interested in creating a simulation that mirrors the real-world in the sense that given a set of routes (generated via UO-MMR or SO-MMR), the users execute their routes and the effect is observed. In this sense, we use MATSim as a one-shot simulator to test the viability and performance of the different UO-MMR and SO-MMR generated routes. 

\subsection{Simulation setup}
\label{sec:sim-setup}

Let user $i$ be defined by a tuple $(t_i,o_i,d_i)$, where $(o_i,d_i)$ is its origin-destination pair and $t_i$ is the departure time of the user from its origin $o_i$. There are two types of population in our simulation: (i) population $X$ which contains users that have the UO-MMR generated paths and (ii) population $Y$ which contains users that have the SO-MMR generated paths. As previously mentioned, the social ratio $\alpha$ along with the total population size $P$ determines the size of the two populations (i.e.~$|X|=(1-\alpha) P$ and $|Y|=\alpha P$. In our simulations, we vary the social between zero and one (i.e.~$\alpha\in[0,1]$) with increments of $0.1$. This allows us to assess the penetration level of our platform and the routes it suggests (assuming participants accept the suggested routes) required to have an appreciable impact on performance. 

Prior to the start of the simulation, users do not have routes assigned to them. Instead, as each user is slotted to enter the network, say at time $t_i$, they are given UO-MRR or SO-MMR routes, depending on their population assignment (i.e.~$X$ for UO-MMR and $Y$ for SO-MMR), generated using the current state of the simulator. In the case of the SO-MMR, generated routes are a function of historical route assignment (from the beginning of the simulation) and current network conditions.  


We define the state to be the volume of users $\mu_e$ for each edge $e\in E$. For a given time $k$, if $\delta$ users  enter the network, then edge-specific costs are determined by
the following BPR \cite{manual1964bureau} model:
\[T(\mu_e+\delta_e)=t_{f,e}\left(1+a_e\frac{\mu_e+\delta_e}{c_e}\right)^{4}\]
where $\sum_e\delta_e=\delta$ and $\delta_e$ is the number of additional users assigned to edge $e$.

\begin{algorithm}[htbp]
	\KwData{Total Demand, List of Agent Itineraries}
	\KwResult{Travel time, Congestion statistics of all simulations}
    
    \Begin{
    
    	$\alpha \longleftarrow 0.0$\;
		$P \longleftarrow$ Total Demand\;
    
        \Repeat{$\alpha > 1.0$}{	
            Start the simulation\;
			Initialize network with $\alpha P$ agents from $X$ and $(1-\alpha)P$ agents from $Y$\;
			Get state of network $x_e$\;

            \ForEach{$i \in X+Y $}{
               		
                    \uIf{agent $i \in X$}{
                        $P_{i}^{u} \longleftarrow$ Compute UO-MMR path for agent $i$ using $x_e$\;
                        Drop agent $i$ in $P_{i}^{u}$\;

                     }
                     \uElseIf{agent $i \in Y$}{
                          $P_{i}^{s} \longleftarrow$ SO-MMR path for agent $i$ using $x_e$\;
                          Drop agent $i$ in $P_{i}^{s}$\;

                    }
                    
             }
                 
               \ForEach{$i \in X+Z $}{
               		\If{agent $i$ is at $d_i$ (destination)}{
        				\textbf{Stop Simulation}\;
      				}
                }
 				
                Save Travel time, Congestion statistics of simulation\;
            
               $k \longleftarrow \alpha + 0.1$\;
      }
    }
    \Return{Travel-time, congestion statistics of all simulations}
    \label{algo-sim_execution}
	\caption{Simulation execution}
\end{algorithm}

\begin{figure}[t]
\begin{center}
\centerline{\includegraphics[width=.75\columnwidth]{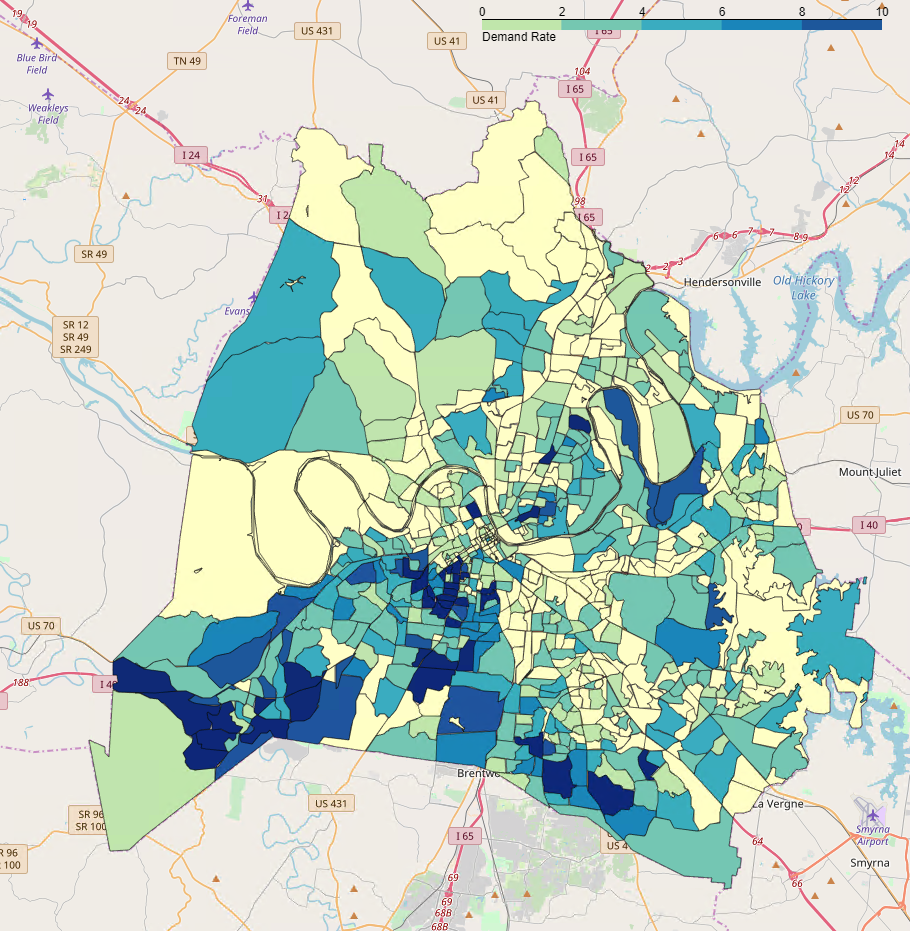}}
\caption{Vanderbilt Employees Demand distribution}
\label{fig:demand_dist}
\end{center}
\vspace{-0.3in}
\end{figure}

\subsection{Simulation Results}
\label{subsec:sim-results}

We have used Vanderbilt employees trip distribution data for modeling agents used in our simulation. and \textbf{Traffic analysis zone (TAZ)} dataset from US Census \cite{census2016nashville} to model demand distribution of Vanderbilt employees in each TAZ. TAZ is a special area delineated by state and/or local transportation officials for tabulating traffic-related data, especially journey-to-work and place-of-work statistics. Based on the demand distribution in each TAZ, we have sampled random points in each TAZ based on likelihood. So, if one TAZ ha 100 people and another TAZ has 1000 people, then the likelihood of picking people from 1000 people TAZ will be 10 times compared to TAZ of 100 people. Figure~\ref{fig:demand_dist} shows the demand rate of Vanderbilt employees in each TAZ.

In our Vanderbilt dataset, we have job type for each employees. Time for each employee (agent in simulation) is randomly assigned based on their job types as shown in Table~\ref{table:datasets}.  So, simulation is done for peak hour traffic at 7.30-10am for onward journey and 4pm-7pm for return journey.

\begin{table}[t]
\centering
\begin{tabular}{|p{1.5cm}|p{2.8cm}|p{2.5cm}|} 
\hline
Job type & Onward Journey & Return Journey\\
\hline
Faculty & 7.30am-9am & 5pm-7pm\\
\hline
Students & 8am-10am & 4pm-6pm\\
\hline
Staff & 7am-8am & 5pm-6pm\\
\hline
\end{tabular}
\vspace{0.02in}
 \caption{Temporal distribution of Vanderbilt employees. }
 \label{table:datasets}
\vspace{-0.3in}
\end{table}

We have collected historical traffic data for Nashville from January 1 to January 31 and have used Random Forest regression \cite{natingga2017data} to build a traffic speed prediction model for a typical weekday (Ignoring weekend data). This model takes total capacity, free flow speed, number of lanes of the link and hour of the day to predict the traffic speed.

The simulation is executed as described in Section \ref{sec:sim-setup}. We use average travel time (from their origins to destinations) across all users in both $X$ and $Y$, considered together, as a performance metric for the system (of course other performance metrics can and should be considered as appropriate). Since the best travel-times across roads are different, we normalize the results:
\[\text{normalized travel-time} = \frac{\text{(actual travel-time) - (best travel-time)}}{\text{best travel-time}}\] 

\begin{figure}
\begin{subfigure}[h]{0.48\linewidth}
\includegraphics[width=1.0\columnwidth]{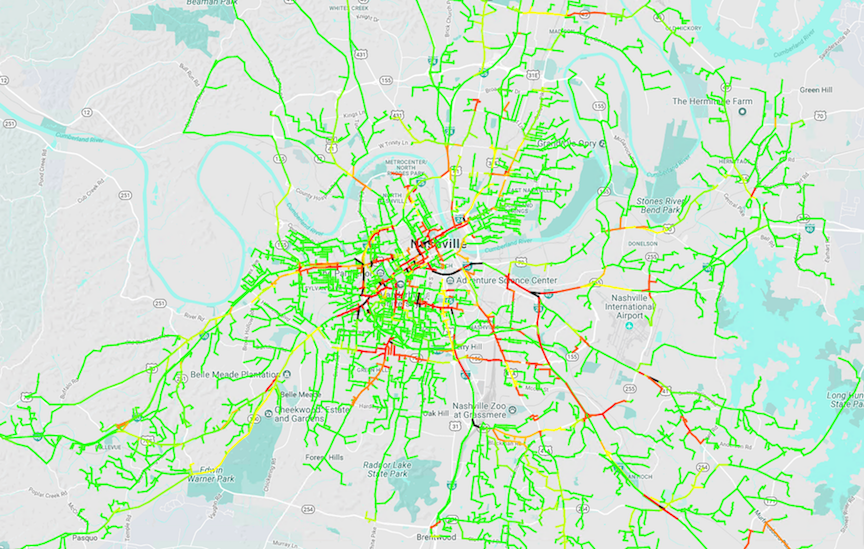}
\caption{Congestion heatmap for $\alpha=0.0$ simulation}
\end{subfigure}
\hfill
\begin{subfigure}[h]{0.48\linewidth}
\includegraphics[width=1.0\columnwidth]{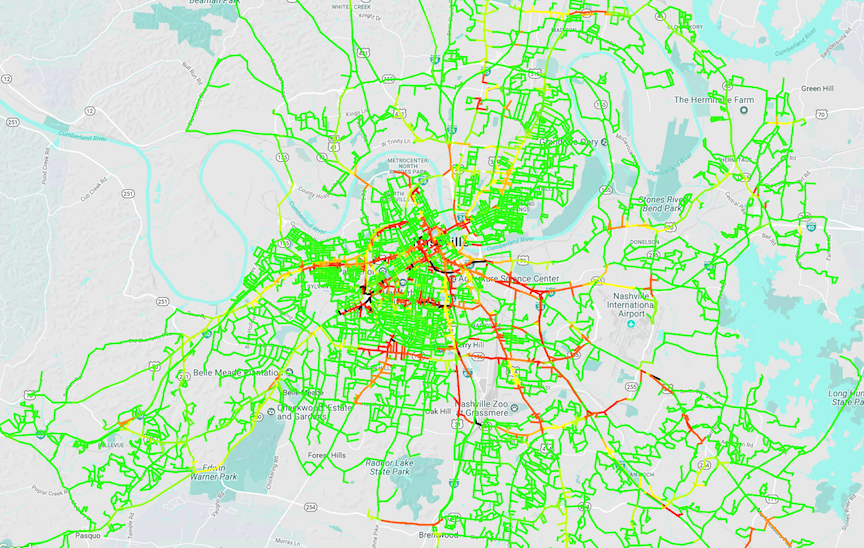}
\caption{Congestion heatmap for $\alpha=1.0$ simulation}
\end{subfigure}

\caption{This figure shows the congestion heatmaps of the links used by agents during simulation;(a) Congestion heatmap for $\alpha=0.0$ simulation when all agents are following UO-MMR; (b) Congestion heatmap for $\alpha=1.0$ simulation when all agents are following SO-MMR.}
\label{fig:simulation-congestion}
\vspace{-0.1in}
\end{figure}

Figure~\ref{fig:simulation-congestion} shows the change in congestion heatmaps of the links used by agents between the simulation having no agents following SO-MMR suggestions (i.e.~$\alpha=0.0$) and the simulation where all the agents follow SO-MMR suggestions (i.e.~$\alpha=1.0$). This heatmap contains congestion both from MATSim agents and background traffic. As the heatmap shows, agents in social ratio 1.0 prefer alternative routes and modes than agents for social ratio 0.0. Since the heatmap contain background traffic too, there are links which are heavily congested in both the heatmaps and our SO-MMR router plays no role in decreasing congestion in those links.

\begin{figure}[tphb]
\begin{center}
\centerline{\includegraphics[width=1\columnwidth]{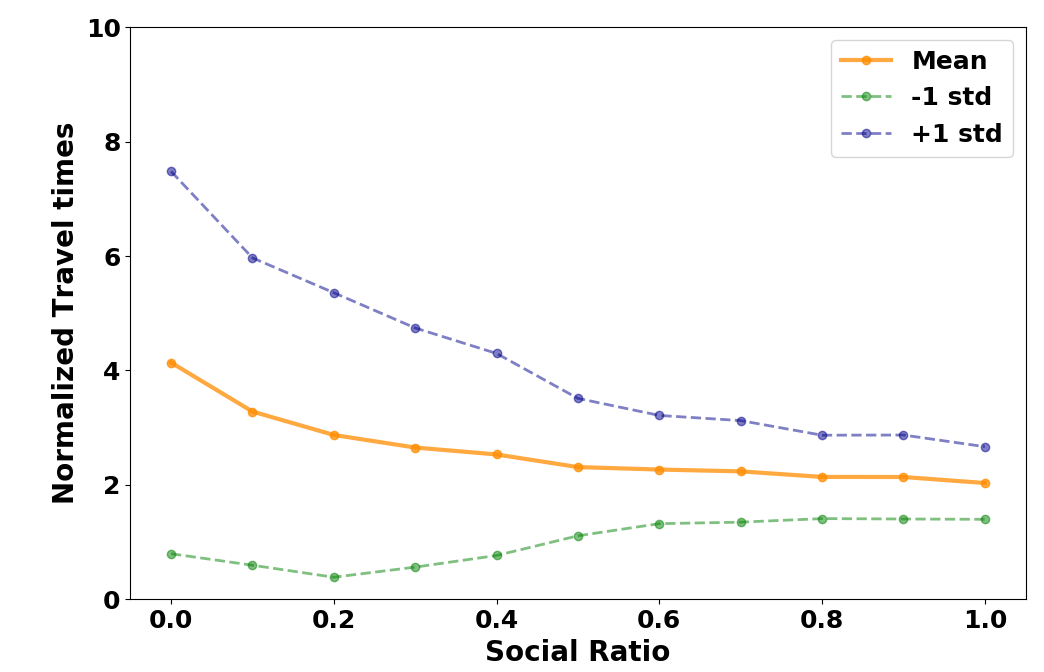}}
\caption{Mean and variance of the normalized travel-times. As the number of agents using SO-MMR increases, the average travel-time of all agents in the system decreases as does the variation in travel-times.}
\label{fig-mean-variance}
\end{center}
\end{figure}

Figure~\ref{fig-mean-variance} shows the normalized travel times of all the agents in simulation. The figure shows that as the number of users using SO-MMR increases, the average travel time of all agents in the system decreases as does the variation in travel-times. This implies that as more agents follow SO-MMR suggestions, congestion in the system decreases as measured by the average travel-time in the network. This decrease in travel times is because agents are provided alternate routes with car, bus and walk. After social ratio of 0.7, the mean remains almost same, while the variance continues to decrease.

\begin{figure}[tphb]
\begin{center}
\centerline{\includegraphics[width=1\columnwidth]{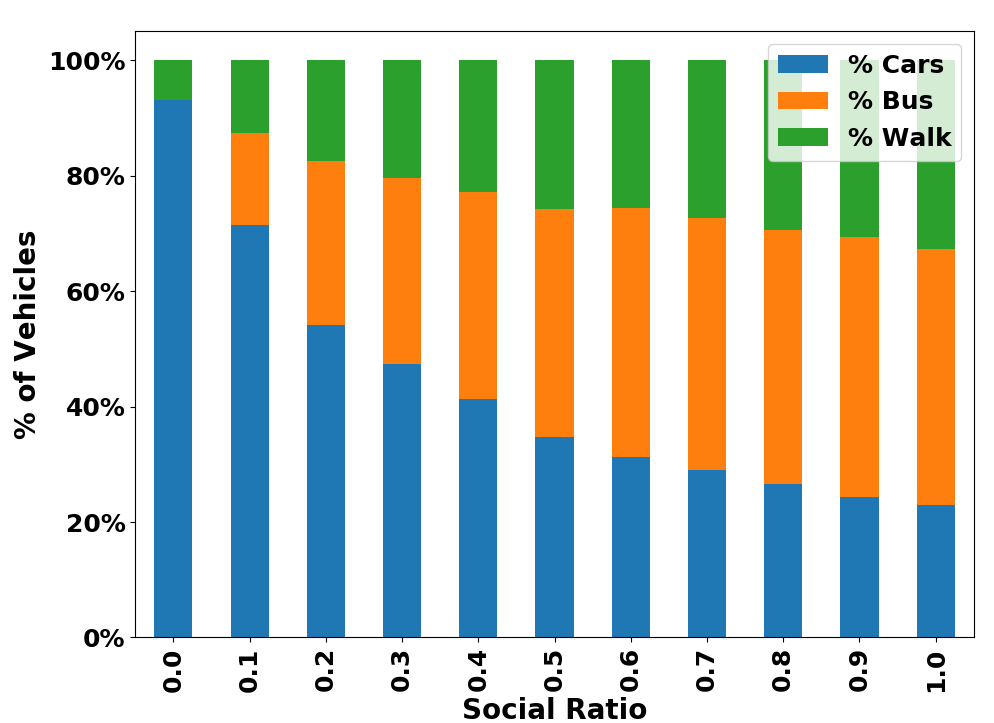}}
\caption{Mode distribution for different ratios of users using SO-MMR. This figure shows that as ratio of users using SO-MMR increases, the number of agents using transit increases, while at the same time the number of agents using their personal vehicle decreases. It indicates that, as $\alpha$ increases, SO-MMR is routing more agents through transit in order to decrease congestion.} 
\label{fig-mode-dist}
\end{center}
\end{figure}

Figure~\ref{fig-mode-dist} shows the mode distribution for different ratios of agents using SO-MMR. The results imply that as ratio of users using SO-MMR increases, the number of agents using transit increases, while at the same time number of agents using their personal car decreases. That is, as $\alpha$ increases, SO-MMR is routing more agents through transit and walk to decrease congestion. It should be noted that in this plot, agents are using multiple modes to reach their destination. Large increase in transit ridership is due to the fact that agents increasingly use bus only in some legs of their entire trip. Having rental data and parking location data for cars might give a better result.

\begin{figure}[tphb]
\begin{center}
\centerline{\includegraphics[width=1\columnwidth]{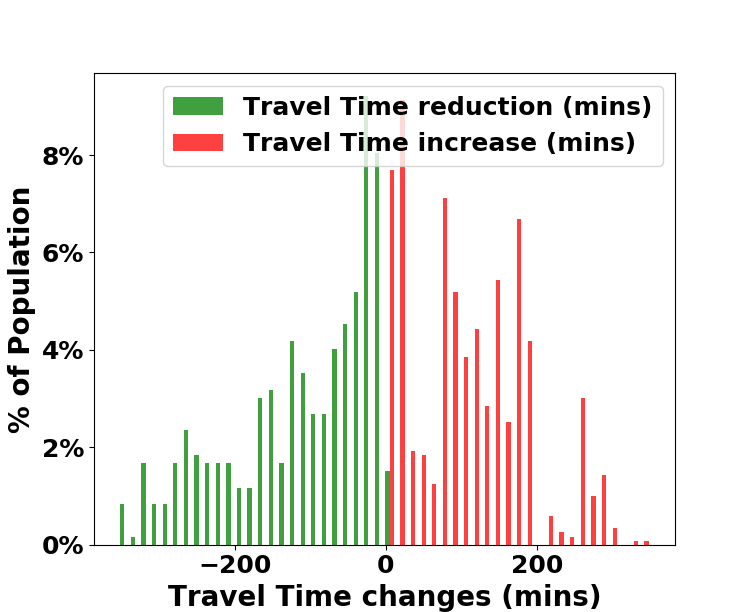}}
\caption{The change in travel-time agents experience between the simulation having no agents following SO-MMR suggestions (i.e.~$\alpha=0.0$) and the simulation where all the agents follow SO-MMR suggestions (i.e.~$\alpha=1.0$). There is a huge variation in this figure, with many agents observe drastic increase in travel time ($200 min$). The rest of the population experiences a significant decrease in travel-time under the SO-MMR suggested routes.}

\label{fig-travel-time-changes}

\end{center}
\end{figure}

Figure~\ref{fig-travel-time-changes} shows the change in travel-time each agent experiences between the simulation having no agents following SO-MMR suggestions (i.e.~$\alpha=0.0$) and the simulation where all the agents follow SO-MMR suggestions (i.e.~$\alpha=1.0$). Total decrease in travel time for some agents outweighs the total increase in travel time of rest of the agents in population. There is a huge variation in this figure, with many agents observe drastic increase in travel time ($200 min$). Such variations are due to increased congestion on roads and limited bus availability (frequency) for many trips. It also depends on the transit network. In Nashville transit network has hub-spoke pattern, where most of the agents have to go to downtown, wait for bus and then go to Vanderbilt on another bus. Having more frequency of buses in some routes and more point-to-point connections can give better results.

\section{Discussion}
\label{sec:discussion}

Our proposed routing algorithm improves on multi-objective A$^\ast$ \cite{tung1992multicriteria} by including congestion information and by suggesting routes to users that are constructed given the plans of other users and are near-optimal routes that can decrease congestion in system. Specifically, even at a low penetration (i.e.~small values of the social ratio $\alpha$), we are able to achieve an improvement in system-level performance.  To see a significant improvement, we are developing techniques that actively increase the social ratio by incentivizing users to participate in our SO-MMR-based platform as well as to adopt the routes it suggests. 

Our simulation results are based on the assumption that users are taking the route proposed to them by SO-MMR. This assumption will however, not hold true in real life situations where users may not be inclined to take a socially optimal route. 
Moreover, as our results indicate,  there are some users who are worse off when using SO-MMR as compared to UO-MMR. Hence we need our routing model to consider the uncertainty of a user actually taking the route, consider fairness in terms of route assignment (as some users may be more likely to get slower routes as a function of other socio-economic characteristics such as income level that are not accounted for in our model), and also propose mechanisms to incentivize users to take a socially optimal route. We are developing a probabilistic model in which we update a prior for the routes different user types adopt. This probabilistic model will serve as a starting point for designing incentives to increase SO-MMR route adoption.

While we were able obtain simulation results that support our arguments, we are also aware of that this is a closed simulation and we did not have any input from the real world. Furthermore, the simulation was conducted with only a few fixed itineraries and we have only considered travel time as a metric for providing user-optimal routes to the agents in simulation. To show more realistic simulation, however, we need to consider common travel demands of users and include fare model for transit, shared vehicles, parking, and other costly services in our simulation.

\section{Conclusions}
\label{sec:conclusion}

We have demonstrated that by taking a proactive approach and leveraging multiple modes in routing decisions, congestion can be decreased. As part of our future work, we will extend our current approach to include (a) incentives for users to take a socially optimal route, (b) modeling uncertainty of users actually following the route proposed by router and (c) realistic simulation that can help us better understand urban vehicular dynamics in a city which potentially can be a tool for city planners to understand the interventions they need to make for encouraging the transition towards a socially optimal multi-modal routing platform.

\section*{Acknowledgments}
This work is sponsored in part by the National Science Foundation under the award number CNS-1646912 (UW), CNS-1647015 (Vanderbilt) and CNS-1528799 (Vanderbilt) and in part by the Vanderbilt Initiative in Smart City Operations and Research, a trans-institutional initiative funded by the Vanderbilt University.

\bibliographystyle{ACM-Reference-Format}
\bibliography{papers}

\end{document}